\newif\iffigs\figstrue
\documentclass[12pt]{article}
\usepackage{latexsym,amssymb}
  \iffigs
   \input{epsf}
\else
\fi

\newcommand{\sect}[1]{\setcounter{equation}{0}\section{#1}}

\newcommand{\eq}{\begin{equation}}
\newcommand{\eqa}{\begin{eqnarray}}
\newcommand{\en}{\end{equation}}
\newcommand{\ena}{\end{eqnarray}}
\newcommand{\enn}{\nonumber \end{equation}}

\def\sk{\vskip .4cm}
\def\noi{\noindent}

\def\Om{\Omega}
\def\al{\alpha}
\def\la{\lambda}
\def\be{\beta}
\def\ga{\gamma}
\def\Ga{\Gamma}

\def\Cb{\bar{C}}

\def\epsi{\varepsilon}
\def\we{\wedge}

\def\de{\delta}

\def\part{\partial}

\def\Gabo{{\bf \mbox{\boldmath $\Ga$}}}

\def\Cb{{\bf \mbox{\boldmath $C$}}}

\def\n2{{{N+1} \over 2}}

\def\square{{\,\lower0.9pt\vbox{\hrule \hbox{\vrule height 0.2 cm
\hskip 0.2 cm \vrule height 0.2 cm}\hrule}\,}}

\def\Q.E.D.{\rightline{$\Box$}}

\def\unop{1  \! \mbox{I}}

\begin{document}
\begin{titlepage}
\vskip -1cm \rightline{DFTT-14/2000}
\rightline{March 2000} \vskip 1em
\begin{center}
{\large\bf New $AdS_3 \times G/H$ compactifications of chiral IIB
 supergravity }
\\[2em]
L. Castellani ${}^{1,2}$ and L. Sommovigo ${}^1$
\\[.7em] ${}^1${\sl Dipartimento di
Scienze e Tecnologie Avanzate,
 East Piedmont University, Italy;} \\ ${}^2$ {\sl
Dipartimento di Fisica Teorica and Istituto Nazionale di Fisica
Nucleare\\ Via P. Giuria 1, 10125 Torino, Italy.} \\
{\small castellani@to.infn.it, sommovigo@mfn.unipmn.it}\\[2em]
\end{center}
\vskip 4 cm
\begin{abstract}
\sk
We present a new class of solutions of $D=10$, $N=2$ chiral supergravity. A
nonvanishing background for the field strength $G_{MNR}$ of the complex
two-form triggers $AdS_3 \times M_7$ compactifications, where
$M_7$ is a 7-dimensional compact manifold. When $M_7$ is a nonsymmetric
coset space $G/H$, we can always find a set of constants
$G_{MNR}$ covariantly conserved and thus satisfying the field equations.
For example the structure constants of $G$ with indices in the $G/H$ directions
are a covariantly conserved tensor.  In some symmetric $G/H$, where these structure
constants vanish, there may still exist conserved 3-forms, yielding
$AdS_3 \times G/H$  solutions.

The conditions for supersymmetry of the $AdS_3 \times M_7$
compactifications are derived, and tested for the  supersymmetric
solutions $AdS_3 \times S^3 \times T^4$ and
$AdS_3 \times S^3 \times S^3 \times S^1$.

Finally, we show that the $AdS_3 \times S^3 \times T^4$ supersymmetric background
can be seen as the $\sigma = 0$ limit
 in a one-parameter class of solutions of the form
$AdS_3 \times S^3 \times \mathbb{C}P^2$,
 the parameter $\sigma$ being
 the inverse ``radius" of
 $\mathbb{C}P^2$. For $\sigma \not= 0$ all supersymmetries are
 broken.

\end{abstract}

\vskip 2cm
 \hrule
  \vskip .2cm
   \noi {\footnotesize
  Supported in part by   EEC  under TMR contract
 ERBFMRX-CT96-0045}

\end{titlepage}
\newpage
\setcounter{page}{1}

\sect{Introduction}

Spontaneous anti de-Sitter compactifications of $D=10$ chiral $N=2$ supergravity
have received some attention recently, in the light of the conjectured
 AdS/CFT correspondence \cite{adscft}. In particular $AdS_5 \times M_5$
 solutions \cite{romans,pw}, where $M_5$ = compact space, have been used to glean
 information on the related 4-dimensional conformal field theory
 \cite{adscft,ads5,pw,pw2}.
 The presence of a (complex) three-form field strength can trigger
 compactifications on $AdS_3 \times M_7$, an example being the
 known solutions $M_7 = S^3 \times T^4$ and $M_7 = S^3 \times
 S^3 \times S^1$, whose corresponding two dimensional conformal theory
 has been investigated in
 \cite{s3t4cft}. Further ref.s on IIB compactifications on spheres
 can be found in \cite{ads3s}.

 In this Letter we present a general class of $AdS_3 \times M_7$ solutions with
 $M_7$ = compact 7-dimensional coset space $G/H$, and derive the
 conditions for these solutions to be supersymmetric backgrounds.
 This class of compactifications may be relevant for the study of the
 two-dimensional CFT's examined in ref.s \cite{ads3goverh}.

The theory contains a complex anti-Weyl gravitino $\psi_{M}$ and a
complex Weyl spinor $\lambda$. The bosonic fields are: the
graviton $g_{MN}$, a complex antisymmetric tensor $A_{MN}$, a real
antisymmetric tensor $A_{MNRS}$ (restricted by a self-duality
condition) and a complex scalar $\phi$. There is a global $U(1)$
symmetry that rotates the two supersymmetry charges into each
other. According to a general mechanism in supergravity theories,
the scalars can be interpreted as coordinates of noncompact coset
spaces. Here the complex scalar $\phi$ parametrizes the coset
$SU(1,1)/U(1)$.

After setting the spinor fields to zero, the field equations read
\cite{js2b,hw,cp}:
 \eqa
 & &2 R_{MN}=P_M P_N^* + P_M^* P_N + {1 \over 6}
 F^{PQRS}_{~~~~~~M} F_{PQRSN} + \nonumber \\
 & & ~~~~~~~~~~~~~~+{1\over 8} ( G^{PQ}_{~~~M} G^*_{PQN}
 + G^{*~PQ}_{~~~~~M} G_{PQN} - {1\over 6} g_{MN} G^{PQR}
 G^*_{PQR}) \label{einsteineq}\\
 & & F_{M_1-M_5}={1 \over 5!} \epsilon_{M_1-M_5N_1-N_5}
 F^{N_1-N_5} \label{Feq}\\
 & & (\nabla^S - i Q^S)G_{MNS}=P^S G^*_{MNS}-{2i\over 3} F_{MNPQR}
 G^{PQR} \label{Geq}\\
 & & (\nabla^M - 2i Q^M) P_M= - {1 \over 24} G^{PQR} G_{PQR}
 \label{scalareq}
 \ena
where  $R_{MN} \equiv R^S_{~M~SN}$ and the curvature two-form is
defined as $R^S_{~M} \equiv dB^S_{~M}+B^S_{~N}\we B^N_{~M}$; the
vectorial quantities $P_M$ (complex) and $Q_M$ (real) are related
to the scalar fields (and derivatives thereof), $F_{PQRSN}$ and
$G_{PQN}$ to the field strengths of the four-form and of the
two-form, and $\nabla$ is the Lorenz covariant derivative. Here we
have adopted the normalizations and conventions of  \cite{js2b};
note however a sign correction in (\ref{scalareq}), already found
in \cite{cp} and noted also in \cite{pw2}. Moreover the following
Bianchi identities hold (a consequence of the field definitions):
\eqa & & (\nabla_{[M} - 2iQ_{[M})P_{N]}=0,~~~\partial_{[M} Q_{N]}=
-i P_{[M} P^*_{N]}\label{bianchipq}\\ & &(\nabla_{[M} -i
Q_{[M})G_{NRS]}=-P_{[M} G^*_{NRS]}\label{bianchig}\\ &
&\partial_{[N} F_{M_1..M_5]}={1 \over 8} \mbox{Im} ~G_{[NM_1M_2}
G_{M_3M_4M_5]}^* \label{bianchif} \ena

 The supersymmetry variations of the
bosonic fields are proportional to Fermi fields, and  these vanish
in the type of backgrounds we are considering. On the other hand,
the supersymmetry variations of the fermionic fields are:
 \eqa
 & & \de \la = i \Ga^M \epsi^* P_M - {1\over 24} i G_{MNP} \Ga^{MNP}
 \epsi  \\
 & & \de \psi_M = (\nabla_M - {i\over 2} Q_M) \epsi +
 {i \over 480} F_{N_1-N_5}\Ga^{N_1-N_5} \Ga_M \epsi  + \nonumber\\
 & &~~~~~+ {1 \over 96} (\Ga_M^{~N_1-N_3} G_{N_1-N_3} - 9
 \Ga^{N_1N_2} G_{MN_1N_2}) \epsi^*
 \ena
(in backgrounds with $\psi=0, \la=0$) \cite{js2b}. A solution is
 supersymmetric if there exist spinors $\epsi$ for which
these variations vanish.

\sect{The Ansatz for the $AdS_3 \times M_7$ solutions}

We use the index conventions
 \sk
 M,N,P...= 1-10

  m,n,p...= 1-3 (run on $AdS_3$)

   a,b,c...= 4-10 (run on  $M_7$).
   \sk
\noi and the flat ``mostly minus" D=10 metric $\eta =
(+,-,-;-,-,-,-,-,-,-)$. With the Ansatz:
 \eqa
 & & g_{MN}=\mbox{metric of }AdS_3 \times M_7 \nonumber\\
 & &\mbox{ fermions} = 0;~~P_M=Q_M=F_{M_1-M_5}=0 \nonumber \\
 & &G_{mnp}=e \epsilon_{mnp},~~~G_{abc}=g J_{abc} \label{ansatz1}
 \ena

\noi where $e$ and $g$ are complex constants and $J_{abc}$ a real
constant antisymmetric tensor, the field eq.s (\ref{scalareq})
take the form:
 \eq
 6e^2-g^2 J^2=0, ~~J^2 \equiv J_{abc} J_{abc} >0 \label{scalareq2}
 \en
implying
 \eq
 g=\rho {\sqrt{6}\over J} e,~~\rho = \pm 1 \label{gerelation}
 \en
 Substituting the Ansatz(\ref{ansatz1})and the relation
 (\ref{gerelation}) into the remaining field eq.s yields:
 \eqa
 & &  R_{mn}={1\over 4} |e|^2 g_{mn} \label{adsricci}\\
 & &  R_{ab}={3\over 4} { J_{ab}\over J^2}~|e|^2,~~J_{ab}
 \equiv J_{acd}J_{bcd} \label{M7ricci}\\
 & & \nabla^c J_{abc}=0,~~~~\nabla=M_7-\mbox{covariant derivative}
 \label{nablaj}
  \ena
Eq. (\ref{Feq}) is trivially satisfied, while eq. (\ref{Geq}) with
free indices $m,n$ holds because $\epsilon_{mnp}$ is a covariantly
conserved tensor in $AdS_3$. Moreover it is immediate to check that
the Bianchi identities (\ref{bianchipq})-(\ref{bianchif}) are satisfied.
 Thus our Ansatz is a solution of the
classical 2b supergravity equations provided eq.s (\ref{adsricci})
- (\ref{nablaj}) hold. The first equation fixes the $AdS_3$
radius. If there exist an $M_7$ - covariantly conserved
three-index antisymmetric tensor $J_{abc}$ in $M_7$ the third
equation is satisfied, and the second equation becomes a condition
on the Ricci tensor of $M_7$.

As we show in the following, such  $J_{abc}$ always exist
 in nonsymmetric coset spaces $G/H$ ( and in various symmetric
 $G/H$).
Moreover $J_{ab}$ is diagonal, allowing in most cases $G/H$ to
solve eq. (\ref{M7ricci}) after a vielbein rescaling.


\sect{$G/H$ geometry and the tensor $J_{abc}$}


The structure constants of
$\mathbb{G}=\mathbb{H}+ \mathbb{K}$  are defined by
 \eqa
 & &[H_i,H_j]=C_{ij}^{~~k} H_k ~~~~~~~~~~~~~~~~~~H_i \in \mathbb{H}\nonumber\\
 & &[H_i,K_a]=C_{ia}^{~~j} H_j + C_{ia}^{~~b}K_b~~~~~K_a \in \mathbb{K}
 \nonumber\\
 & &[K_a,K_b]=C_{ab}^{~~j}H_j+C_{ab}^{~~c}K_c
 \ena
where the index conventions are obvious.
As discussed in ref. \cite{pvntrieste} (p. 251), whenever $H$ is
compact or semisimple one can always find a basis
of $K_a$ such that the structure constants $C_{ia}^{~~j}$ vanish.
In that case the $\mathbb{G}=\mathbb{H}+\mathbb{K}$ split, or equivalently the coset space $G/H$
is said to be {\sl reductive}. For this reason we will deal in this
paper only with reductive coset spaces. Another important observation
is that when $G/H$ is reductive the structure constants $C_{ia}^{~~b}$
can always be made antisymmetric in $a,b$ by an appropriate
redefinition $K_a \rightarrow N_a^{~b} K_b$ \cite{pvntrieste,pvnleshouches}.
\sk
For
later use we recall the expression of the $G/H$ Riemannian connection
\eq
 B^a_{~b}= {1\over 2}\left(- {r_b r_c\over r_a} C_{bc}^{~~a} +
  {r_a r_c\over r_b}
 \eta_{bg} C_{dc}^{~~g} \eta^{ad} + {r_a r_b \over r_c} \eta_{cg}
 C_{db}^{~~g} \eta^{ad}\right) V^c - C_{bi}^{~~a} \Omega^i
 \label{GHconnection}
\en
where $V^c$ and $\Om^i$ are the $K$ and $H$ vielbeins
respectively, and we have allowed for rescalings $r_c$ of $V^c$ in
the isotropy irreducible subspaces of $K$, see ref.s
\cite{CRWcosets,CDFbook,LCGH}. These subspaces correspond to the block-diagonal
pieces of the matrices $C_{bi}^{~~a}$, so that
\eq
{r_a \over
r_b}~C_{ia}^{~~b}= C_{ia}^{~~b} \label{irred}
\en
The $G/H$ Riemann curvature is defined by
$R^{a}_{~b} \equiv dB^{a}_{~b} + B^a_{~c} \we B^{c}_{~b} \equiv
R^a_{~b~de} V^d \we V^e$ and reads \cite{LCGH}:
 \eq
 R^a_{~b~de} = {1\over 4} {r_d r_e \over r_c} \Cb_{bc}^{~~a} C_{de}^{~~c}
 + {1\over 2} r_d r_e~C_{bi}^{~~a} C_{de}^{~~i} + {1\over 8}
 \Cb_{cd}^{~~a} \Cb_{be}^{~~c} - {1 \over 8} \Cb_{ce}^{~~a}
 \Cb_{bd}^{~~c} \label{GHcurvature}
 \en
with
 \eq
 \Cb_{bc}^{~~a} \equiv {r_b r_c\over r_a}~ C_{bc}^{~~a} - {r_a r_c\over r_b}~
 C_{ac}^{~~b} - {r_a r_b \over r_c}~ C_{ab}^{~~c}
 \en

Consider now the field eq. (\ref{nablaj}), i.e.:
 \eq
 B_{e~~[a}^{~d} J_{bc]d}~ \eta^{ec}=0 \label{Jtest}
 \en
\noi where the connection 1-form components are defined by
 $B^d_{~a} \equiv B_{c~~a}^{~d} V^c$.
 One possible choice for a $J_{abc}$  satisfying (\ref{Jtest})
 is given by:
 \eq
  J_{abc}=C_{abc} \equiv C_{ab}^{~~G} \ga_{cG},~~\mbox{G runs on
the group $G$, $\ga$ = Killing metric} \label{JeqC}
\en
Indeed $B_{e~~[a}^{~d} C_{bc]d}~ \eta^{ec}=0$ holds for the
following reason. Observe that the left hand side is an
$H$-invariant tensor, since the connection components, the
structure constants $C_{abc}$ and the Killing metric are
 all $H$-invariant tensors. By $H$-invariant tensor we mean, for
 example, that:
 \eq
 \de B_{c~b}^{~a} \equiv C_{ic}^{~~~d}
B_{d~b}^{~a}- C_{id}^{~~~a} B_{c~b}^{~d}+ C_{ib}^{~~~d}
B_{c~d}^{~a}=0
\en
  i.e. the adjoint action of $H$ on $B$ vanishes.
 It is not difficult to prove that in (\ref{GHconnection})
 the term multiplying
$V^c$ is $H$-invariant. In fact each of the three terms
within parentheses in (\ref{GHconnection}) is $H$-invariant,
as one can show by using Jacobi identities and (\ref{irred}).

In general the only $H$-invariant tensor with two free indices is
the Killing metric (except in the special case of $S^2$, where one
has also $\epsilon_{ab}$), and therefore $B_{e~~[a}^{~d} C_{bc]d}~
\eta^{ec}$, being antisymmetric in its free indices, has to
vanish. \sk In conclusion, the choice (\ref{JeqC}) satisfies the
field eq.s (\ref{nablaj}). Moreover $J_{ab} \equiv J_{acd}
J_{bcd}$, being a symmetric $H$-invariant tensor, must be
proportional to the Killing metric in the $H$-isotropy subspaces.
\sk
  The structure constants $C_{abc}$ are not the most general
  solution to eq. (\ref{Jtest}). For example also Antisymm ($C_{ab}^{~~c}$),
  i.e. the antisymmetrization of $C_{ab}^{~~c}$ on its three indices,
  satisfies eq. (\ref{Jtest}).  In fact there may exist $J_{abc}$
  tensors satisfying (\ref{Jtest}) even for symmetric $G/H$;
  this happens obviously for $S^3=SO(4)/SO(3)$, where $J_{abc}$ is
  proportional to $\epsilon_{abc}$, or less trivially
   in the case $G/H =S^3 \times \mathbb{C}P^2$
  discussed in Section 7.


\sect{Supersymmetry conditions}


We adopt the following real representation of the $D=3+7$ gamma
matrices:
\eq
\Gabo
_M = (\ga_m \otimes \unop_{8 \times 8}\otimes \sigma_2,~~\unop_{2 \times 2}
\otimes \Ga_a \otimes \sigma_1 )
\en
where the $SO(1,2)$ gamma matrices are:
\eq
\ga_1=\pmatrix{0 & -i \cr
              i & 0  \cr},~~
\ga_2=\pmatrix{0 & i \cr
              i & 0  \cr},~~
\ga_3=\pmatrix{ i& 0 \cr
               0 & -i \cr}
\en
and the real $SO(7)$ gamma matrices are given by the octonion
structure constants (totally antisymmetric):
\eqa
& &(\Ga_a)_{bc}=a_{abc},~~(\Ga_a)_{b8}=\de_{ab}\\
& & a_{abc}:~~[123]=[165]=[257]=[354]=[367]
=[246]=[147]=1
\ena
The supersymmetry parameter $\epsi$ has the same Weyl chirality
as the gravitino $\psi$, i.e. $\Gabo_{11} \epsi=-\epsi$ (anti-Weyl).
Any anti-Weyl $SO(1,9)$ spinor can be decomposed as :
\eq
\epsi = c_N \xi^N \otimes \eta^N \otimes \left( \matrix{0\cr 1\cr}
\right)
\en
where $\xi^N$ and $\eta^N$ are real $SO(1,2)$ and $SO(7)$ spinors,
respectively, and $c_N \in \mathbb{C}$. Substituting in the
$\lambda$ supersymmetry condition: \eq G_{MNR} \Gabo^{MNR} \epsi=0
\Rightarrow (e \epsilon_{mnr} \Gabo^{mnr}+ g J_{abc} \Gabo^{abc})
\epsi=0
\en
yields:
 \eq
  c_N \xi^N \otimes (6 e \unop + g J_{abc} \Ga^{abc})
 \eta^N=0,
 \en
which can hold only if there exist $SO(7)$ spinors satisfying:
 \eq
( \unop + \rho { J_{abc}\over \sqrt{6} J} \Ga^{abc}) \eta=0
\label{condlambda}
\en
where we have also used (\ref{gerelation}).

 Consider now the
$\psi$ supersymmetry condition:
 \eqa
 & & (d  + {1\over 4} B^{mn} \Gabo_{mn} + {1 \over 4} B^{ab} \Gabo_{ab} )
 \epsi + \nonumber \\
 & & ~~~~~+{1 \over 96} V^m (G_{abc} \Gabo_m \Gabo^{abc} - 9 e \epsilon_{mnr}
  \Gabo^{nr}) \epsi^* + \nonumber \\
 & & ~~~~~+ {1 \over 96} V^d (G_{abc} \Gabo_d^{~abc} + e \epsilon_{mnr} \Gabo_d
  \Gabo^{mnr} - 9 G_{dab} \Gabo^{ab} ) \epsi^* =0 \label{condpsi0}
  \ena
  $V^m$ and $V^d$ are the $AdS_3$ and $M_7$ vielbeins respectively.
  Substituting the $\Gabo$-matrix and $\epsi$ decomposition leads to:
  \eqa
  & & c (\part_m \xi + {1\over 4} B^{rs}_{~~m} \ga_{rs} \xi) - {1 \over 4} i c^*
  e \ga_m \xi =0 \label{condpsi1}\\
  & & c(\part_c \eta + {1\over 4} B^{ab}_{~~c} \Ga_{ab} \eta) - {1\over 8}
  c^* G_{cab} \Ga^{ab} \eta =0 \label{condpsi2}
  \ena
  where we have dropped the index $N$ since the conditions can be satisfied
  only if they  hold separately for every $N$. Moreover use of
  (\ref{condlambda}) has been necessary in order to achieve the
  factorization of (\ref{condpsi0}) into (\ref{condpsi1}) and (\ref{condpsi2}).
  The modulus of $c$ is irrelevant in (\ref{condpsi1}) and (\ref{condpsi2}),
  and we can set $c=\exp(i\varphi)$. Then the integrability condition for
  (\ref{condpsi1}) is:
  \eq
  [{1\over 4} R^{mn}_{~~rs} \ga_{mn} - ({1\over 4})^2 \exp(-4i\varphi) e^2
  \ga_{rs} ] \xi = 0 \label{integrability1}
  \en
  The field equations (\ref{adsricci}) tell us that the $AdS_3$ curvature is:
  \eq
  R^{mn}_{~~rs}= {1\over 4} |e|^2 \de^{mn}_{rs}
  \en
so that the integrability condition can be satisfied if and only
if the phase of $c$ is such that:
 \eq
  |e|^2 = e^2 \exp(-4i\varphi)
\Rightarrow e= \al |e| \exp(2i\varphi) ,~~~ \al = \pm 1
\label{ephi}
\en
cf. ref. \cite{igor}.
Using this relation and (\ref{gerelation}) into (\ref{condpsi2})
yields finally:
 \eqa
 & & \part_m \xi + {1\over 4} B^{rs}_{~~m} \ga_{rs} \xi - {1 \over 4} i
 \al |e| \ga_m \xi =0 \label{condpsi11}\\
& &\part_c \eta + {1\over 4} B^{ab}_{~~c} \Ga_{ab} \eta -
{\sqrt{6}\over 8}
  \al \rho |e| {J_{cab}\over J} \Ga^{ab} \eta =0 \label{condpsi22}
  \ena
  The integrability condition for (\ref{condpsi22}) reads:
  \eq
  (R^{ab}_{~~cd} + {3 \over 2} |e|^2 {J_{cae} J_{dbe} \over J^2} )
   \Ga_{ab} \eta
  =0 \label{integrability2}
  \en
  The $AdS_3 \times M_7$ solution preserves $N$ supersymmetries if and only
  if there exist $N$ independent $SO(7)$ spinors $\eta$ satisfying simultaneously eq.s
  (\ref{condlambda}) and (\ref{condpsi22}).

 In next Section we test our formulae in the case of the two known
 supersymmetric solutions, corresponding to $G/H = S^3 \times T^4$
 and
 $G/H = S^3 \times S^3 \times S^1$.

 \sect{The supersymmetric solutions $AdS_3 \times S^3 \times T^4$ and
 $AdS_3 \times S^3 \times S^3 \times S^1 $}

We'll treat the two solutions simultaneously. The $J_{abc}$
tensors are simply the $\epsilon_{abc}$ Levi-Civita tensors in the
$S^3$ directions, so that Einstein field equations (\ref{M7ricci})
are respectively:
 \eqa
 & & R_{ab}={1 \over 4} |e|^2 \de_{ab},~~a,b=1,2,3 \nonumber\\
 & & R_{ab}=0,~~a,b=4,5,6,7 \label{ricciS3T4}
 \ena
 and
 \eqa
 & & R_{ab}={1 \over 8} |e|^2 \de_{ab},~~a,b=1,2,..6 \nonumber\\
 & & R_{ab}=0,~~a,b=7  \label{ricciS3S3S1}
 \ena
fixing the radii of the $S^3$ spheres.

In the real gamma matrix representation of Section 4, the $\de
\la=0$ supersymmetry condition is satisfied by any linear
combination of the four independent spinors
$\eta_1,\eta_2,\eta_3,\eta_8$, the 8-dimensional spinor $\eta_a$
having the a-th component as only nonvanishing component. This
holds for both solutions. On the other hand, the $\de \psi=0$
supersymmetry condition is satisfied by 8 independent real spinors
$\eta_1^{\pm},\eta_2^{\pm},\eta_3^{\pm},\eta_8^{\pm}$, the
$^{\pm}$ referring to the sign $\al$ of (\ref{condpsi22}). These
spinors depend on the $M_7$ coordinates. Again this holds for both
solutions. The two compactifications have then $N=8$
supersymmetries, or 16 real conserved supercharges (since $\xi$
has two real components).

\sect{Other $AdS_3 \times G/H$ solutions}

A complete classification of all $AdS_3 \times G/H$ solutions based on the
Ansatz (\ref{ansatz1}) is postponed to a later publication.
This requires to find the most general $J_{abc}$ that solves eq.
(\ref{Jtest}) for each 7-dimensional $G/H$.

Here we give some selected examples, choosing
some particular $J$ tensors. In fact, all the 7-dimensional
$G/H$ cosets classified in \cite{CRWclassification} are solutions of the
IIB field equations, after suitable rescalings of the coset
vielbeins. But the $G/H$ list of IIB solutions is actually larger than
the one relevant for $D=11$ supergravity compactified on
$AdS_4 \times G/H$. Indeed in the IIB case the field equations
do not force $G/H$ to be an Einstein space, but rather to be
``isotropy Einstein", i.e. with a Ricci tensor $R_{ab}$ proportional to
$\de_{ab}$ in each isotropy irreducible subspace of $G/H$.
 The proportionality constant can also vanish: when this happens in
 one-dimensional subspaces
$S^1$ factors are allowed (they were excluded in the list of
\cite{CRWclassification}).

\subsection{$AdS_3 \times M^{pqr}$}

The $M^{pqr}$ spaces have been studied in detail in ref.s
\cite{CDFMpqr, CDFbook,fre1}. They have three isotropy irreducible
subspaces, allowing three independent rescalings $a,b,c$ of the
vielbeins corresponding to the (1,2), 3, (4,5,6,7) directions and
preserving the $SU(3) \times SU(2) \times U(1)$ isometry. Their
Ricci tensor is given by \cite{CDFMpqr}: \eqa & & R_{mn} = {1\over
4}b^2(2-{b^2\over c^2} q^2) \de_{mn},~~m,n=1,2 \nonumber\\ & &
R_{33}= {9 a^4 p^2 + 2 b^4 q^2 \over 8 c^2} \nonumber \\ & &
R_{AB}= {3 \over 16} a^2 (4-3 {a^2 \over c^2} p^2)
\de_{AB},~~A,B=4,5,6,7 \label{ricciMpqr1} \ena For $q \not= 0$, $p
\not=0$ we can redefine: \eq a= {q\over p} \ga \sqrt{{2\al\over
3}},~b=\ga \sqrt{2 \be},~~c=q \ga
\en
so that the Ricci tensor becomes:
\eqa
& & R_{mn} = \ga^2 \be (1-\be) \de_{mn},~~m,n=1,2 \nonumber\\
& & R_{33}= \ga^2 (\be^2 + {1\over 2} {q^2 \over p^2} \al^2) \nonumber \\
& & R_{AB}= {1\over 2} \ga^2 \al (1-{1\over 2} \al) {q^2\over p^2}
\de_{AB} \label{ricciMpqr2}
\ena
An explicit check reveals that taking $J_{abc}$ to be the Levi-Civita tensor
in the directions 1,2,3 (as in the case of the $S^3 \times T^4$
solution) and
otherwise zero
satisfies the condition (\ref{Jtest}). Then the field equations are
as in (\ref{ricciS3T4}), and are satisfied by the Ricci tensor in (\ref{ricciMpqr2})
when the rescalings are:
\eq
\al=2,~~ \be= {1 \pm \sqrt{1-16 {q^2 \over p^2}}\over
4},~~\ga= {1 \over 4 \be (1-\be)} |e|^2
\en
requiring $p/q \geq 4$. The particular case $q=0$, corresponding to
$S^2 \times S^5$, is also a solution: the rescalings are then:
\eq
a^2 = {1\over 6} |e|^2,~~b^2 = {1\over 2} |e|^2,~~c^2= {1\over 8} p^2 |e|^2
\en

The $\de \la$ supersymmetry condition is satisfied by the same
spinors $\eta_1,\eta_2,\eta_3,\eta_8$ discussed in the previous
Section. However these spinors do not satisfy the $\de \psi$
condition,
and therefore these solutions are not supersymmetric.

There are other possible choices for
$J_{abc}$. For example $J_{abc}=$ Antisymm($C_{ab}^{~~c}$) leads to field
equations that can still be solved by a set of different
rescalings $\al, \be, \ga$. In this case the four directions 4,5,6,7
are not Ricci flat.

\subsection{$AdS_3 \times N^{010}$}

The $N^{010}$ coset spaces are a special case in the class of
$N^{pqr}$ coset spaces studied in ref.s
\cite{CRNpqr,LCNpqr,LCGH,fre2} in the context of $D=11$
supergravity compactifications. They can be realized as the
quotient: \eq
  N^{010}={SU(3) \times SU(2) \over
SU(2) \times U(1)} \label{N010}
\en
where the $SU(2)$ in the denominator is diagonally embedded in
$G = SU(3) \times SU(2)$. In this formulation the full isometry
of $N^{010}$ comes from the left action of $G$
 \cite{CRNpqr,LCGH}. The $N^{010}$ geometry has been
 studied in detail in  \cite{LCGH}, and the Ricci tensor
 is given by:
  \eqa
& & R_{ab}=\left(\al^2+{1\over 32}{\be^4\over  \al^2}\right)
\de_{ab} \label{ricci1N010}\\
 & & R_{AB}={3 \over 4} \be^2 \left( 1 -{1 \over 16} {\be^2 \over
 \al^2} \right) \de_{AB} \label{ricci2N010}
 \ena
Again one can check explicitly that the same $J_{abc}=
\epsilon_{abc}, a,b,c=1,2,3$ used in the previous solutions
satisfies the field equations (\ref{Jtest}). Then $AdS_3 \times N^{010}$
is a solution provided the rescalings are fixed to the values:
\eq
\al = \pm {1\over 6} |e|,~~\be = \pm {2 \over 3} |e|
\en
As in the previous case supersymmetry is absent because (\ref{condpsi22}) has no
solutions.

In a similar way we easily find that for every 7-dimensional coset space $G/H$
of the classification \cite{CRWclassification} there exist a set
of constants $J_{abc}$ such that $AdS_3 \times G/H$  is a solution
for IIB supergravity, with the exception of the round 7-sphere
$SO(8)/SO(7)$.
Also, every $G/H \times S^1$ space
with $G/H$ = nonsymmetric 6-dimensional coset is a solution of the IIB equations
for appropriate rescalings and $J_{abc}= C_{abc}$.

What remains to be done is to determine in each case the most general
$J_{abc}$, and check whether there are instances in which
both supersymmetry conditions can be satisfied simultaneously,
yielding supersymmetric solutions as in  Section 5.

We expect that, in order to make contact with the $d=2$
superconformal theories discussed in \cite{ads3goverh}, we need to
extend our Ansatz to a nonconstant scalar field.

\sect{Asymptotic supersymmetry: $S^3 \times (\mathbb{C}P^2 \rightarrow
T^4)$}

We study here a particular class of solutions , characterized by a
continuous parameter $\sigma$. Consider the cosets $G/H=M^{010}$, a
special case of the $M^{pqr}$ class with $S^3 \times \mathbb{C}P^2$
topology \cite{CDFMpqr}.
In general a rescaling of the $G$ structure constants given by:
\eq
C_{G_1G_2}^{~~~G_3} \longrightarrow {r_{G_1} r_{G_2}\over
r_{G_3}} C_{G_1G_2}^{~~~G_3}
\en
still defines a Lie algebra. Take all $r$ = 1 except those in the
$4,5,6,7$ coset directions, for which $r=\sigma$. Then the Ricci
tensor becomes:
\eqa
& & R_{mn} = {1\over 4}b^2(2-{b^2\over c^2} ) \de_{mn},~~m,n=1,2 \nonumber\\
& & R_{33}= { b^4 \over 4 c^2} \nonumber \\
& & R_{AB}= {3 \over 4} a^2 \sigma^2  \de_{AB},~~A,B=4,5,6,7
\label{ricciM010}
\ena
The limit $\sigma \rightarrow 0$ corresponds to a group contraction
yielding $S^3 \times T^4$: in fact it amounts to sending the radius of
$\mathbb{C}P^2$ to infinity. For any $\sigma$, i.e. for any
$\mathbb{C}P^2$-radius, $AdS_3 \times S^3 \times \mathbb{C}P^2$ is a
solution of the IIB equations if we choose the $J_{abc}$ tensor to
be:
\eq
 J_{123}=1,~~J_{345}=J_{367}=\sigma
 \en
 and the rescalings:
 \eq
 a^2= {|e|^2 \over 3( 1+2 \sigma^2)},~~b^2=\pm c |e|,~~c =
 {\sigma^2 +1 \over 2\sigma^2 +1} |e|
 \en
 One can see easily that the $\de \lambda$ supersymmetry condition is
 satisfied by just one of these solutions, corresponding to
 $\sigma=0$, i.e. the $N=8$ supersymmetric $AdS_3 \times S^3 \times T^4$
 compactification. All the other values of $\sigma$ break
 supersymmetry. Thus we obtain a class of continuously connected
 solutions, parametrized by $\sigma$, all of them nonsupersymmetric
 except in the limit $\sigma=0$. For $\sigma \not= 0$, $S^3$ is a
 squashed three-sphere, cf. eq.s (\ref{ricciM010}).

\vskip 1cm

{\bf Acknowledgements} \sk It is a pleasure to acknowledge useful
discussions with Pietro Fr\'e and Igor Pesando.


\vfill\eject
\end{document}